# Optimized Passive Optical Networks with Cascaded-AWGRs for Data Centers

Mohammed Alharthi
*School of Electronic and Electrical Engineering*
*University of Leeds*
Leeds, United Kingdom
elmaalh@leeds.ac.uk

Sanaa H. Mohamed
*School of Electronic and Electrical Engineering*
*University of Leeds*
Leeds, United Kingdom
s.h.h.mohamed@leeds.ac.uk

Barzan Yosuf
*School of Electronic and Electrical Engineering*
*University of Leeds*
Leeds, United Kingdom
b.a.yosuf@leeds.ac.uk

Taisir E. H. El-Gorashi
*School of Electronic and Electrical Engineering*
*University of Leeds*
Leeds, United Kingdom
t.e.h.elgorashi @ksu.edu.sa

Jaafar M. H. Elmirghani
*School of Electronic and Electrical Engineering*
*University of Leeds*
Leeds, United Kingdom
j.m.h.elmirghani@leeds.ac.uk

***Abstract*—The use of Passive Optical Networks (PONs) in modern and future data centers can provide energy efficiency, high capacity, low cost, scalability, and elasticity. This paper introduces a passive optical network design with 2-tier cascaded Arrayed Waveguide Grating Routers (AWGRs) to connect groups of racks (i.e. cells) within a data center. This design employs a Software-Defined Networking (SDN) controller to manage the routing and assignment of the networking resource while introducing multiple paths between any two cells to improve routing, load balancing and resilience. We provide benchmarking results for the power consumption to compare the energy efficiency of this design to state-of-the-art data centers. The results indicate that the cascaded AWGRs architecture can achieve up to 43% saving in the networking power consumption compared to Fat-Tree data center architecture.**

## I. INTRODUCTION

Current data centers have many limitations such as high cost, high latency, limited scalability, management complexity and low throughput [1], [2]. Therefore, it is becoming necessary to design new data center architectures that resolve these limitations. Previous studies indicated interest in designing energy efficient access and core communication networks [3]-[11] and also energy-efficient solutions for data center designs [12]-[17]. Traditional data center network designs, for example Fat-Tree and VL2 in [18], use switches and routers that offer high performance while on the other hand they are associated with high power consumption and maintenance costs. The studies on energy efficient data centers have advanced in three directions which are the use of commodity switches, all-optical data centers, or hybrid switching [19].

Commodity switch-based architectures, which are divided into switch-centric and server-centric, have much lower cost and power consumption compared to enterprise-level switches [20]. In designing a commodity switch-based architecture, some of the factors to consider include the bandwidth capacity, latency, complexity, scalability, resilience, and cost. Optical networking technologies are promising to enhance the performance of data centres as well as to reduce energy consumption and latency, in addition to increasing the data rates and flexibility [21]. Hence, they are considered as solutions to the shortcomings of electronic switching data centers.

The use of Passive Optical Network (PON) technology, which has a confirmed high performance in access networks, in modern and future data centers can provide energy efficiency, high capacity, low cost, scalability, and elasticity in their designs. Various PON technologies were used in data center networks while keeping the Top-of-the-Rack (ToR) electronic switches. The PON technologies in data center networks include Arrayed Waveguide Grating Routers (AWGRs), Orthogonal Frequency Division Multiplexing (OFDM) PON, and Wavelength Division Multiplexing (WDM) PON [22]-[25]. In [26], five novel PON-based data center architectures were proposed to offer high capacity for intra and inter rack interconnections at low cost, and high energy-efficiency in future Data Center Networks (DCNs). These architectures, compared with current DCN architectures, replace access, aggregation and core electronic switches with different passive intra-rack and inter-rack interconnections in addition to an Optical Line Terminal (OLT).

The third novel architecture was discussed in [12], [27]. This design contained two AWGRs to provide full interconnection within the PON cell that contains four PON groups (i.e., four racks). Each server in this architecture is equipped with an array of photodetectors and tunable lasers for wavelength detection and transmission [12]. Optimizing the wavelength routing and assignment for inter-rack communication was addressed in [12]. Moreover, the architecture was compared to current data centers (i.e., Fat-Tree and BCube) in terms of energy efficiency and the results showed that the proposed architecture achieved power savings of up to 45% compared to the Fat-Tree design and up to 80% compared to the BCube design.

In [28], a Passive Optical Interconnect (POI) design was introduced for data centers based on cascade wavelength routing components (i.e. arrayed waveguide gratings (AWGs)) [28]. Data center architectures based on AWG were studied in terms of achieving non-blocking routing, efficient wavelength assignment, and supporting multicast traffic in [29]. Software-Defined Networking (SDN) can provide coordination and flexibility in wavelength assignment and routing in passive optical networks [30]. SDN was also proposed in [31] to control routing in data centers to facilitate the channel access for inter-cell communication [27],[31].

We benefit from our previous contributions that utilized mixed integer linear programming (MILP) optimization to tackle a range of problems such as processing placement and caching in IoT/Edge networks [32]-[35], greening core and data center networks [36]-[41], use of machine learning and network optimization in data centre healthcare systems [42]-[45] and the use of network coding to improve the energy efficiency of core networks [46], [47].

In this paper, we propose a design for a PON-based data centers with 2-tier cascaded AWGRs for inter-cell communication that provides multipath routing and increases the scalability. This architecture uses four OLT switches instead of a single OLT switch to reduce the oversubscription ratio and enhance the load balancing of traffic compared to the work in [12]. In this design, each cell is connected to two different AWGRs in the first layer to achieve two uplinks and two different AWGRs in the first layer to achieve two downlinks. The first layer of AWGRs connects the cells and the second layer of AWGRs while the second layer of AWGRs connects the first layer of AWGRs with the four OLT switches. This architecture utilizes SDN to assign suitable wavelength and paths for communication between servers in different cells and control the traffic flow through various OLT switches. To achieve all-to-all non-blocking interconnection between the cells and the OLT switches, we develop and utilize a MILP model to optimize and hence maximize the connectivity for the proposed passive network with 2-tier cascaded AWGRs.

The reminder of this paper is organized as follows: Section II describes the proposed architecture which uses a passive optical network with 2-tier cascaded AWGRs for inter-cell communication in data centers. Section III explains the proposed MILP model for optimizing the routing and wavelength assignment in the proposed data center design. Section IV provides the connectivity and wavelength assignment results while Section V presents the power consumption benchmark study to compare our proposed architecture with the Fat-tree architecture. Finally, Section VII provides the conclusions and future work.

## II. The Design of a PON with 2-Tier Cascaded AWGRs for Inter-cell Communication in Data Centers

Figure 1 shows an example of the passive optical network with 2-tier cascaded AWGRs design for inter-cell communication. In this work, we consider 4 cells each with two racks. Each cell connects to the first layer of the AWGRs that includes four AWGRs and each AWGR has eight outputs and inputs to support multipath routing and provide connection between the cells and the second layer of AWGRs. The second layer (i.e., level two) of AWGRs as well includes four AWGRs and each AWGR has eight outputs and inputs to provide connection between the first layer of AWGRs and the four OLT switches. Each input port of an AWGR sends a different wavelength to a different output port and each output port receives from each input port a different wavelength. This can be achieved for the wavelengths routing through cyclic and acyclic designs [48].

Each cell has two links for uplink and two links for downlink. Each uplink connects to a different AWGR in the first layer and each downlink connects to a different AWGR in the first layer (e.g. AWGR_Level_1 in Figure 1). Each AWGR in level 1 has one link to connect to each AWGR in Level 2. Similarly, each AWGR in Level 2 has links to connect to each AWGR at Level 1. Finally, the two layers of AWGRs connect the PON cells to four OLT switches.

The SDN controller enables any cell to connect to any other cell or OLT switch via the 2-tier cascaded AWGRs. It first assigns a suitable wavelength for the servers within each cell. The SDN controller then manages and controls the assignment of routes and wavelength resources for the demands between cells. The SDN controller also manages the reconfiguration of PON protocol relying on the ratio of servers activity in different racks by resetting the transceivers of servers to balance the load through various OLT ports. Therefore, this architecture reduces the oversubscription by allowing servers to connect to different OLT switches.

The number of wavelengths needed for all-to-all inter cell communications is equal to *N* in the case of not considering intra cell communications, where *N* is the number of cells and OLT switches [49]. If intra rack communications is considered, the number of wavelengths is equal to *N*-1 [49]. Therefore, the architecture depicted in Figure 1 requires *N* -1 wavelengths for single path and 2(*N* - 1) for assigning two paths between each communicating pairs (i.e. multipath routing). For example, this design contains four cells and four OLT switches, hence, *N* = 8. Consequently, the total number of wavelengths in this design are equal to (*N* – 1) = 7 wavelengths for single path and 2(8 - 1) = 14 wavelengths for multipath routing. A passive polymer optical backplane is considered for intra rack communication [50].

## III. MILP Model for Optimising the Passive Network with 2-Tier Cascaded AWGRs for Inter-cell Communication of Data Center Design

In this section, we briefly describe the Mixed Integer Linear Programming (MILP) model we developed to optimize the wavelength assignment and routing through the 2-tier cascaded AWGRs to maximize the connectivity for inter-cell communication in the proposed data center architecture shown in Figure 1. The 2-tier cascaded AWGRs PON in the design in Figure 1 supports the connection between cells and the OLT as well as the connections between the cells. The notations that were used to define the sets and parameters in the optimization model are as below:

Sets and parameters:

| | |
|---|---|
| $N$ | Set of all nodes (AWGRs' ports, PON groups (i.e. cells and the OLT) |
| $G$ | Set of PON groups, where $P \subset N$ |
| $W$ | Set of wavelengths |
| $K$ | Set of AWGRs (i.e. 8 AWGRs here) |
| $I_k$ | Set of input ports of AWGR k; $k \epsilon K$ |
| $O_k$ | Set of output ports of AWGR k; $k \epsilon K$ |
| $N_m$ | Set of neighbouring nodes of node $m$; $m \epsilon N$ |

Variables:

| | |
|---|---|
| $\alpha_{sd}^{j}$ | Is a variable that is equal to 1 ($\alpha_{sd}^{j}$= 1) if wavelength $j$; $j \in W$, is used for the connection between a source $s$ and a destination $d$; $s, d \in G$, and otherwise is equal to zero. |
| $\beta_{sd}^{jmn}$ | Is a variable that is equal to 1 ($\beta_{sd}^{jmn}$= 1) if wavelength ; $j \in W$ on link $(m, n)$; $(m, n) \in N$ is |
| | used for a connection $(s,d)$; $(s,d) \in G$ otherwise $\beta_{sd}^{jmn} = 0$. |

entering into a node at a specific wavelength leaves the node at the same wavelength for all the nodes except for the

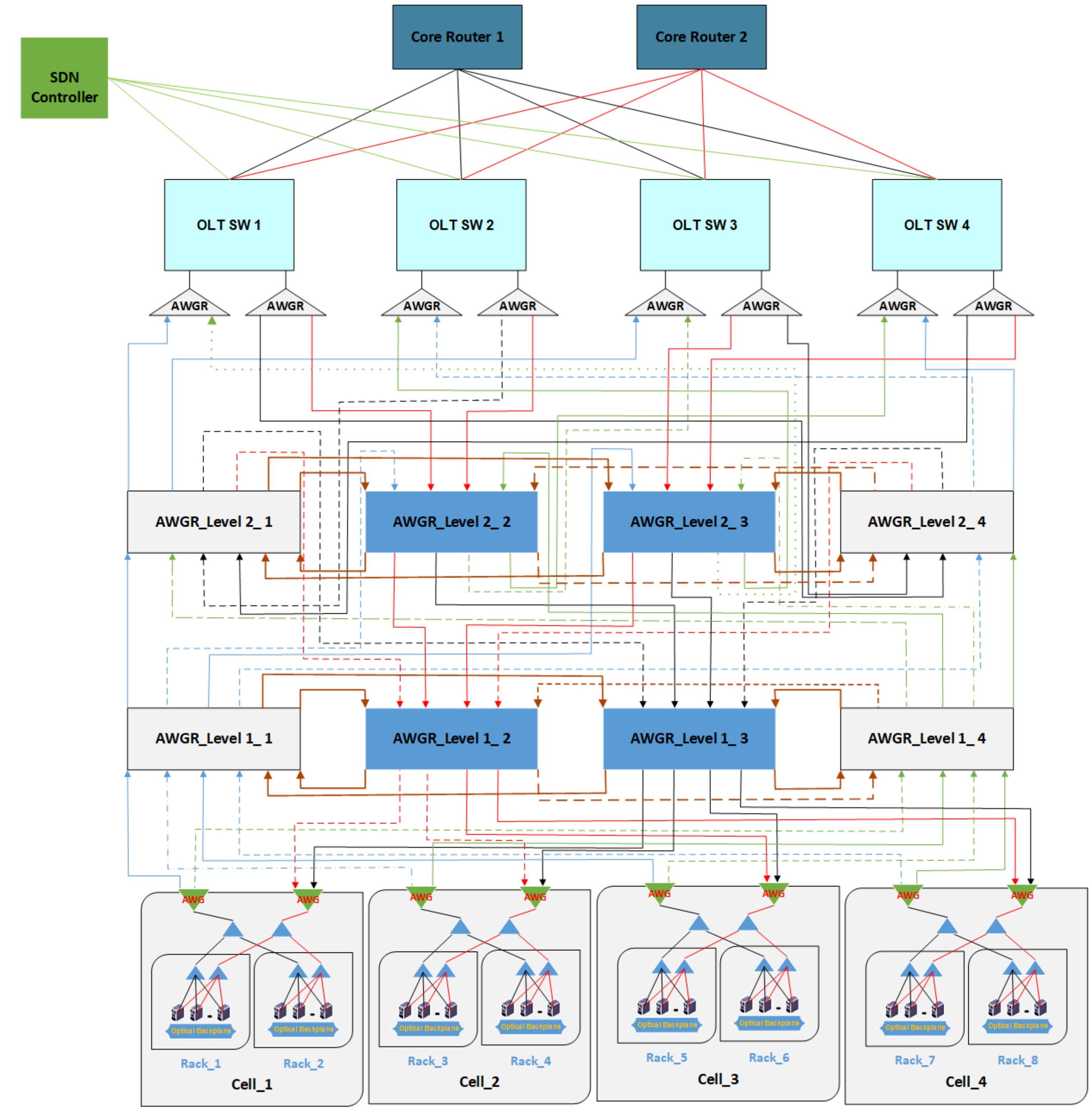

Fig. 1: Passive optical network with 2-tier cascaded AWGRs for inter-cell communication in the proposed data center design.

The model objective is comprised of maximising the total number of connections supported by the AWGRs between PON groups. This objective can be expressed as follows:

*Maximize:*

$$\sum_{s \in G} \sum_{\substack{d \in G \\ s \neq d}} \sum_{j \in W} \alpha_{sd}^{j}. \qquad (1)$$

The model uses wavelength allocation constraints to ensure that each communication between PON groups and OLT and among the PON groups uses two different wavelengths. It also uses constraints to ensure that each destination receives two different wavelengths from different source nodes and each source receives two different wavelengths from different destination nodes. A constraint for the wavelength continuity is used to guarantee the flow destination and source nodes. A routing constraint is also used to guarantee that a wavelength is only used once to connect a pair of source and destination nodes on a physical link. In addition, constraints to ensure correct routing for the AWGRs are considered, where for example the traffic should only be directed from an input port to an output port of an AWGR and not vice versa. The connection between the racks and level 1 of AWGRs is determined by 1:*N* Arrayed Waveguide Grating (AWGs) multiplexers in the uplink and *N*:1 AWGs in the downlink.

## IV. WAVELENGTH ASSIGNMENT RESULTS

Table I illustrates the MILP model results for the AWGR wavelength routing interconnection and the wavelengths the servers use for communication within the PON cell, where, λx is the wavelength number and AWGR_Lx_x represents the AWGR location (i.e. Lx is the level number of AWGR (i.e. Level 1 and Level 2) and x is the number of AWGRs in the same level (i.e. 1, 2, 3 and 4 for each level).) . The order of the AWGR_Lx_x in Table 1 represents the routing through the 2-tier cascaded AWGRs, This design considers Wavelength Routing Network (WRN) with ($N$+1) elements where $N$ is the number of cells and OLT switches that communicate with each other. A WRN requires $N^2$ fibres with $N$ wavelengths or $N$ fibres with $N^2$ wavelengths for ($N$+1) elements to communicate with $N$ elements [49]. This design uses the 2-tier cascaded AWGRs CLOS topology to represent the $N^2$ fibres and the number of wavelengths is equal to 7 for a single path and 14 for multipath routing. Wavelength assignment for communication between the PON groups is shown in table I. If server A located in cell 1 requires communication with server B located in cell 4, a request control message is sent to the SDN controller to enable the communication. If the SDN controller decides to grant the request, then it replies with control messages to servers A and B. The control messages contain information about the wavelength tuning required for servers A and B to communicate. In addition, the SDN controller may decide to group this communication to a common OLT switch. After that, both servers tune their transceivers to wavelength 6 or 12 as the SDN controller decided. Idle servers by default should be tuned to a wavelength that connects them with one of the OLT switches [12].

TABLE I: MILP BASED RESULTS FOR THE WAVELENGTH ASSIGNMENT AND ROUTING FOR THE PON WITH 2-TIER CASCADED AWGRS FOR THE INTER-CELL COMMUNICATION . EACH PAIR ARE ASSIGNED TWO DIFFERENT WAVELENGTHS FOR THE COMMUNICATION. EACH WAVELENGTH IS ROUTED IN A DIFFERENT PATH THROUGH THE TWO LAYERS OF AWGRS.

| | OLT 1 | OLT 2 | OLT 3 | OLT 4 | Cell 1 | Cell 2 | Cell 3 | Cell 4 |
|---|---|---|---|---|---|---|---|---|
| OLT 1 | | | | | $\lambda_8$<br>AWGR_L2_2<br>AWGR_L1_2<br>$\lambda_9$<br>AWGR_L2_2<br>AWGR_L2_4<br>AWGR_L1_3 | $\lambda_7$<br>AWGR_L2_4<br>AWGR_L2_2<br>AWGR_L1_3<br>$\lambda_{11}$<br>AWGR_L2_4<br>AWGR_L1_2 | $\lambda_4$<br>AWGR_L2_4<br>AWGR_L2_3<br>AWGR_L1_3<br>$\lambda_{14}$<br>AWGR_L2_4<br>AWGR_L1_3 | $\lambda_2$<br>AWGR_L2_2<br>AWGR_L1_3<br>$\lambda_3$<br>AWGR_L2_2<br>AWGR_L2_4<br>AWGR_L1_3 |
| OLT 2 | | | | | $\lambda_3$<br>AWGR_L2_2<br>AWGR_L2_1<br>AWGR_L1_2<br>$\lambda_6$<br>AWGR_L2_2<br>AWGR_L1_3 | $\lambda_5$<br>AWGR_L2_1<br>AWGR_L1_2<br>$\lambda_{13}$<br>AWGR_L2_1<br>AWGR_L2_3<br>AWGR_L1_3 | $\lambda_2$<br>AWGR_L2_2<br>AWGR_L2_4<br>AWGR_L1_2<br>$\lambda_{12}$<br>AWGR_L2_1<br>AWGR_L1_3 | $\lambda_4$<br>AWGR_L2_2<br>AWGR_L1_2<br>$\lambda_8$<br>AWGR_L2_2<br>AWGR_L2_1<br>AWGR_L1_3 |
| OLT 3 | | | | | $\lambda_{12}$<br>AWGR_L2_3<br>AWGR_L1_2<br>$\lambda_{14}$<br>AWGR_L2_4<br>AWGR_L2_3<br>AWGR_L1_2 | $\lambda_4$<br>AWGR_L2_4<br>AWGR_L1_3<br>$\lambda_8$<br>AWGR_L2_3<br>AWGR_L1_2 | $\lambda_3$<br>AWGR_L2_4<br>AWGR_L2_2<br>AWGR_L1_3<br>$\lambda_{11}$<br>AWGR_L2_4<br>AWGR_L2_3<br>AWGR_L1_2 | $\lambda_1$<br>AWGR_L2_3<br>AWGR_L1_3<br>$\lambda_5$<br>AWGR_L2_4<br>AWGR_L1_2 |
| OLT 4 | | | | | $\lambda_2$<br>AWGR_L2_3<br>AWGR_L1_3<br>$\lambda_7$<br>AWGR_L2_3<br>AWGR_L2_4<br>AWGR_L1_2 | $\lambda_1$<br>AWGR_L2_1<br>AWGR_L1_3<br>$\lambda_3$<br>AWGR_L2_1<br>AWGR_L2_2<br>AWGR_L1_2 | $\lambda_5$<br>AWGR_L2_1<br>AWGR_L2_2<br>AWGR_L1_2<br>$\lambda_9$<br>AWGR_L2_1<br>AWGR_L1_2 | $\lambda_{13}$<br>AWGR_L2_3<br>AWGR_L2_1<br>AWGR_L1_3<br>$\lambda_{14}$<br>AWGR_L2_3<br>AWGR_L1_2 |
| Cell 1 | $\lambda_7$<br>AWGR_L1_1<br>AWGR_L2_1<br>$\lambda_{13}$<br>AWGR_L1_4<br>AWGR_L2_4<br>AWGR_L2_3 | $\lambda_1$<br>AWGR_L1_4<br>AWGR_L2_3<br>AWGR_L2_4<br>$\lambda_4$<br>AWGR_L1_1<br>AWGR_L2_3 | $\lambda_3$<br>AWGR_L1_4<br>AWGR_L2_1<br>$\lambda_{14}$<br>AWGR_L1_1<br>AWGR_L2_2 | $\lambda_5$<br>AWGR_L1_1<br>AWGR_L2_4<br>AWGR_L2_2<br>$\lambda_{11}$<br>AWGR_L1_4<br>AWGR_L2_2 | | $\lambda_2$<br>AWGR_L1_1<br>AWGR_L1_3<br>$\lambda_9$<br>AWGR_L1_1<br>AWGR_L1_3 | $\lambda_8$<br>AWGR_L1_1<br>AWGR_L1_2<br>$\lambda_{10}$<br>AWGR_L1_1<br>AWGR_L1_3 | $\lambda_6$<br>AWGR_L1_1<br>AWGR_L1_2<br>$\lambda_{12}$<br>AWGR_L1_4<br>AWGR_L1_2 |
| Cell 2 | $\lambda_5$<br>AWGR_L1_4<br>AWGR_L2_3<br>$\lambda_{12}$<br>AWGR_L1_1<br>AWGR_L2_3<br>AWGR_L2_1 | $\lambda_7$<br>AWGR_L1_4<br>AWGR_L2_1<br>$\lambda_8$<br>AWGR_L1_1<br>AWGR_L2_2<br>AWGR_L2_4 | $\lambda_2$<br>AWGR_L1_4<br>AWGR_L2_2<br>$\lambda_4$<br>AWGR_L1_1<br>AWGR_L2_4<br>AWGR_L2_2 | $\lambda_3$<br>AWGR_L1_4<br>AWGR_L2_4<br>$\lambda_{14}$<br>AWGR_L1_1<br>AWGR_L2_1<br>AWGR_L2_2 | $\lambda_1$<br>AWGR_L1_4<br>AWGR_L1_3<br>$\lambda_{11}$<br>AWGR_L1_4<br>AWGR_L1_3 | | $\lambda_6$<br>AWGR_L1_4<br>AWGR_L1_2<br>$\lambda_{13}$<br>AWGR_L1_1<br>AWGR_L1_2 | $\lambda_9$<br>AWGR_L1_1<br>AWGR_L1_2<br>$\lambda_{10}$<br>AWGR_L1_1<br>AWGR_L1_3 |
| Cell 3 | $\lambda_3$<br>AWGR_L1_1<br>AWGR_L2_1<br>AWGR_L2_3<br>$\lambda_6$<br>AWGR_L1_4<br>AWGR_L2_4<br>AWGR_L2_3 | $\lambda_2$<br>AWGR_L1_4<br>AWGR_L2_3<br>$\lambda_5$<br>AWGR_L1_1<br>AWGR_L2_3<br>AWGR_L2_4 | $\lambda_8$<br>AWGR_L1_4<br>AWGR_L2_1<br>AWGR_L2_2<br>$\lambda_9$<br>AWGR_L1_1<br>AWGR_L2_2<br>AWGR_L2_1 | $\lambda_1$<br>AWGR_L1_1<br>AWGR_L2_4<br>$\lambda_4$<br>AWGR_L1_4<br>AWGR_L2_2<br>AWGR_L2_4 | $\lambda_{10}$<br>AWGR_L1_4<br>AWGR_L1_2<br>$\lambda_{13}$<br>AWGR_L1_1<br>AWGR_L1_3 | $\lambda_{12}$<br>AWGR_L1_4<br>AWGR_L1_3<br>$\lambda_{14}$<br>AWGR_L1_4<br>AWGR_L1_2 | | $\lambda_7$<br>AWGR_L1_4<br>AWGR_L1_3<br>$\lambda_{11}$<br>AWGR_L1_4<br>AWGR_L1_3 |
| Cell 4 | $\lambda_9$<br>AWGR_L1_4<br>AWGR_L2_1<br>$\lambda_{11}$<br>AWGR_L1_1<br>AWGR_L2_3 | $\lambda_3$<br>AWGR_L1_1<br>AWGR_L2_4<br>$\lambda_{14}$<br>AWGR_L1_4<br>AWGR_L2_4 | $\lambda_{12}$<br>AWGR_L1_4<br>AWGR_L2_3<br>AWGR_L2_1<br>$\lambda_{13}$<br>AWGR_L1_1<br>AWGR_L2_1 | $\lambda_2$<br>AWGR_L1_1<br>AWGR_L2_2<br>$\lambda_8$<br>AWGR_L1_4<br>AWGR_L2_2<br>AWGR_L2_4 | $\lambda_4$<br>AWGR_L1_4<br>AWGR_L1_2<br>$\lambda_5$<br>AWGR_L1_1<br>AWGR_L1_3 | $\lambda_6$<br>AWGR_L1_4<br>AWGR_L1_3<br>$\lambda_{10}$<br>AWGR_L1_1<br>AWGR_L1_3 | $\lambda_1$<br>AWGR_L1_1<br>AWGR_L1_3<br>$\lambda_7$<br>AWGR_L1_4<br>AWGR_L1_2 | |

## V. Power consumption benchmark

In this section, we present a benchmark study to compare the proposed design and one of the currently used data center architectures (i.e. Fat-Tree) in terms of power consumption.

### *A. Fat-Tree architecture*

A Fat-Tree topology [2] comprises three layers which are the core, aggregation and access layers. In Figure 2, an example is given of a Fat-Tree topology that includes $k = 4$ pods and each pod contain $k/2 = 2$ access switches and $k/2 = 2$ aggregation switches. Aggregation and access switches are linked as a CLOS network topology [51] in a dual form in each pod. In addition, the switches in each pod are linked in a dual form to all core switches. Fat-Tree topology composes $k^2/4$ core switches and supports $k^3/4$ hosts.

The power consumption of the Fat-Tree architecture is calculated by assuming different configurations of Fat-Tree switching fabrics. We assessed power consumption for designs that includes 48 and 96 pods. All commodity switches in this study are assumed to use Cisco commodity switches that consume 30W per switch port [52]. Servers' transceiver in this study are assumed to consume 3W [53]. Equation (2) calculates the power consumption of a Fat-Tree network architecture ($P_F$):

$$P_F = P_r\left(\frac{k^3}{4}\right) + P_s\left(\frac{k}{2}\right) + P_s\left(\frac{k}{2}\right) + P_s\left(\frac{k^2}{4}\right), \qquad (2)$$

where $P_r$ is the power consumption of server's transceiver, $P_s$ is the power consumption of the switch port and $k$ is the number of pods.

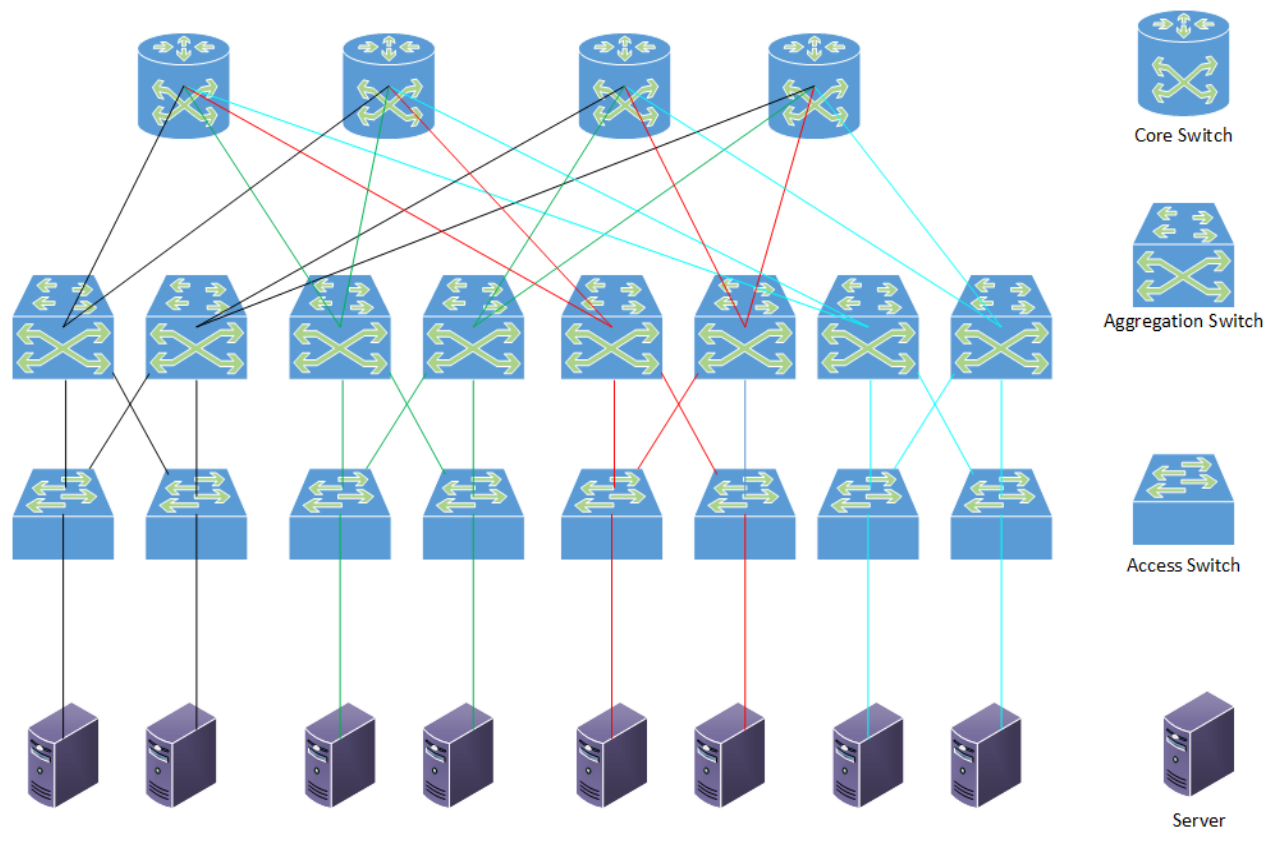

Fig. 2: Fat-Tree topology with k=4.

### *B. The passive network with 2-tier cascaded AWGRs architecture*

The proposed architecture includes four cells which contain servers with wavelength tuneable ONUs that are connected to four OLT switches through 2-tier cascaded AWGRs. The power consumption of this architecture ($P_c$) is calculated as follows:

$$P_c = P_s N_s + P_t N_t + P_U N_U \qquad (3)$$

where $P_s$ is the server transceiver's power consumption and $N_s$ is the number of servers used in the architecture. $P_t$ is the OLT port power consumption, $N_t$ is the number of OLT ports used in the architecture, $P_U$ is the tunable ONU's power consumption and $N_U$ is the number of tunable ONUs used in the architecture.

### *C. Benchmark results*

The benchmark results we considered are for data centers that provide connection for 512, 27,648 and 221,184 servers. The Fat-Tree designs are evaluated with 48 and 96 pods using Cisco's commodity switches [52]. The passive network with cascaded AWGRs for inter-cell communication of data center architecture is composed of 8 racks within 4 PON cells that are connected to four OLTs switches via 2-tier of AWGRs. A cell contains 64 servers in this evaluation and the total number of servers is 512.

The OLT switches in this architecture are assumed to include 16 XG-PON cards and each card has 8 ports. According to [54], the power consumption of an OLT switch is calculated based on the maximum power consumption of the OLT switching, fan card and the power cards.

The power consumption of each element in the Fat-Tree and the proposed design are shown in Table II.

Table III: Power consumption of devices

| Element | Power consumption |
| --- | --- |
| Cisco Catalyst 3850 (WS-C3850-24P) switch | 30 W per port [52] |
| Cisco Catalyst 3850 (WS-C3850-48P) switch | 30 W per port [52] |
| Server's transceiver | 3 W [53] |
| 10-Gb/s OLT port | 14.3 W [54] |
| 10-Gb/s tuneable ONU | 2.5 W [55] |

Figure 3 shows the power savings of the proposed design over the Fat-Tree data center architecture when providing connection for 27,648 and 221,184 servers. The proposed design saves the power consumption by 75.7%, 67.4% and 44.2% compared to the Fat-Tree data center architecture with 512, 27,648 and 221,184 servers, respectively. Using high number of switches for the interconnections plays a pivotal role in the high power consumption in the Fat-Tree data center architecture when compared to the proposed design that replaces the switches by passive devices (i.e. AWGRs and couplers). The power saving accomplished by the proposed design compared to that of the Fat-Tree data center architecture decreases when the total number of server increases. The main reason is that increasing the number of pods (i.e. 48 pods and 96 pods) of Fat-Tree architectures does not increase the power consumption linearly.

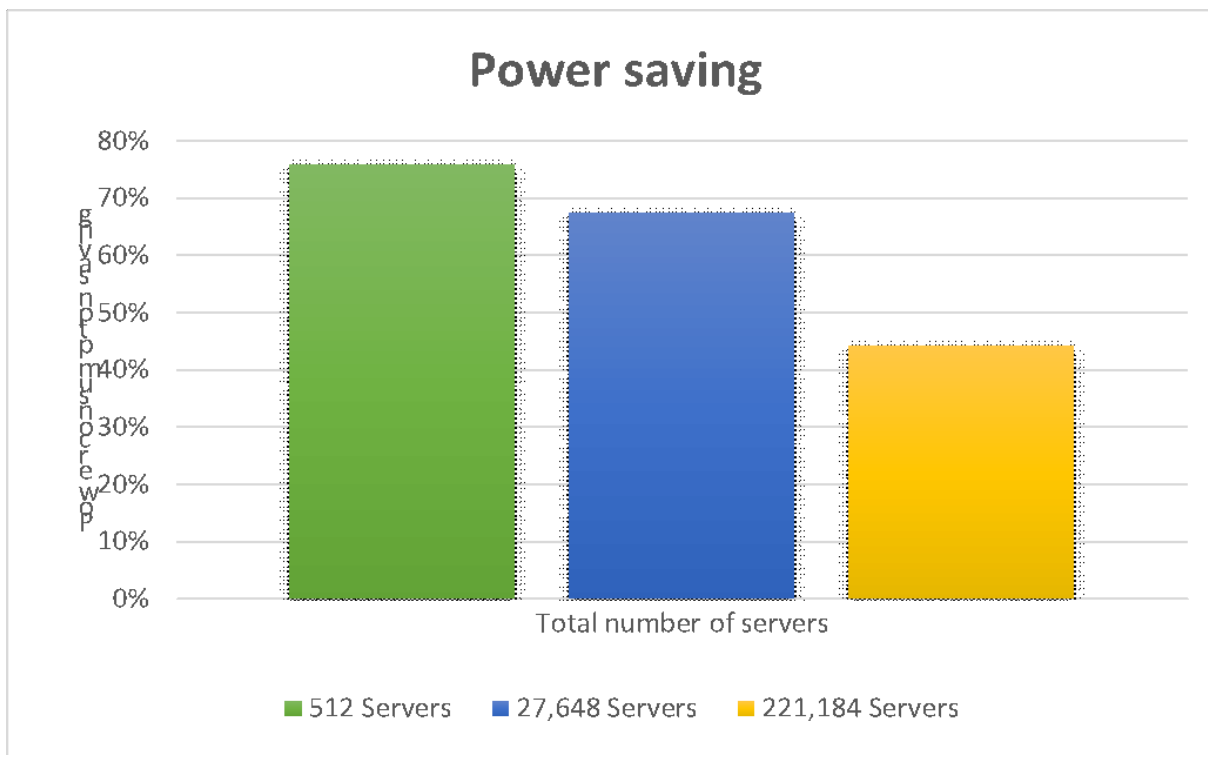

Fig. 3: Benchmark study of power saving comparing Fat-Tree and 2-tier cascade AWGRs architectures.

In addition, the Fat-Tree architecture classification as a switch centric architecture, interconnection between servers and edge switches are designed with single transceivers.

## VI. CONCLUSIONS

This paper proposed a passive optical network with 2-tier cascaded AWGRs for inter-cell communication in data centers. This design eliminates the high power consumption and high cost switches used for edge, aggregation and core switching in current data center architectures. The proposed architecture provides multipath routing and uses four OLT switches instead of using a single OLT which reduces the oversubscription ratios and enhances the load balancing traffic compared to the work in [12]. In addition, a benchmarking study is presented to compare the power consumption of the 2-tier cascaded AWGRs data center architecture to the Fat-Tree data center architecture. The proposed architecture has reduced the power consumption by 75.7%, 67.4% and 44.2% compared to Fat-Tree data center architecture when providing connection for 512, 27,648 and 221,184 servers, respectively.

## ACKNOWLEDGMENT

The authors would like to acknowledge funding from the Engineering and Physical Sciences Research Council (EPSRC), INTERNET (EP/H040536/1), STAR (EP/K016873/1) and TOWS (EP/S016570/1) projects. All data are provided in full in the results section of this paper. The first author would like to thank the Ministry of Interior (MOI), Saudi Arabia for funding his PhD scholarship.

## REFERENCES

[1] S. H. Mohamed, T. E. El-Gorashi, and J. M. H. Elmirghani, "Energy efficiency of server-centric PON data center architecture for fog computing," in 2018 20th International Conference on Transparent Optical Networks (ICTON), 2018, pp. 1-4: IEEE.

[2] A. E. Eltraify, M. O. Musa, and J. M. H. Elmirghani, "TDM/WDM over AWGR based passive optical network data centre architecture," in 2019 21st International Conference on Transparent Optical Networks (ICTON), 2019, pp. 1-5: IEEE.

[3] Z. T. Al-Azez, A. Q. Lawey, T. E. El-Gorashi, and J. M.H. Elmirghani, "Virtualization framework for energy efficient IoT networks," in 2015 IEEE 4th International Conference on Cloud Networking (CloudNet), 2015, pp. 74-77: IEEE.

[4] H. M. M. Ali, A. Q. Lawey, T. E. El-Gorashi, and J. M. H. Elmirghani, "Energy efficient disaggregated servers for future data centers," in 2015 20th European Conference on Networks and Optical Communications-(NOC), 2015, pp. 1-6: IEEE.

[5] X. Dong, T. El-Gorashi, and J. M. H. Elmirghani, "Green IP over WDM networks with data centers," Journal of Lightwave Technology, vol. 29, no. 12, pp. 1861-1880, 2011.

[6] X. Dong, T. El-Gorashi, and J. M. H. Elmirghani, "Energy-efficient IP over WDM networks with data centres," in 2011 13th International Conference on Transparent Optical Networks, 2011, pp. 1-8: IEEE.

[7] X. Dong, T. El-Gorashi, and J. M. H. Elmirghani, "Use of renewable energy in an IP over WDM network with data centres," IET optoelectronics, vol. 6, no. 4, pp. 155-164, 2012.

[8] X. Dong, T. E. El-Gorashi, and J. M. H. Elmirghani, "Joint optimization of power, electricity cost and delay in IP over WDM networks," in 2013 IEEE International Conference on Communications (ICC), 2013, pp. 2370-2375: IEEE.

[9] X. Dong, T. E. El-Gorashi, and J. M. H. Elmirghani, "On the energy efficiency of physical topology design for IP over WDM networks," Journal of Lightwave Technology, vol. 30, no. 12, pp. 1931-1942, 2012.

[10] A. Q. Lawey, T. E. El-Gorashi, and J. M. H. Elmirghani, "Distributed energy efficient clouds over core networks," Journal of Lightwave Technology, vol. 32, no. 7, pp. 1261-1281, 2014.

[11] L. Nonde, T. E. El-Gorashi, and J. M. H. Elmirghani, "Energy efficient virtual network embedding for cloud networks," Journal of Lightwave Technology, vol. 33, no. 9, pp. 1828-1849, 2014.

[12] A. Hammadi, T. E. El-Gorashi, and J. M. H. Elmirghani, "High performance AWGR PONs in data centre networks," in 2015 17th International Conference on Transparent Optical Networks (ICTON), 2015, pp. 1-5: IEEE.

[13] A. Hammadi, T. E. El-Gorashi, M. O. Musa, and J. M. H. Elmirghani, "Server-centric PON data center architecture," in 2016 18th International Conference on Transparent Optical Networks (ICTON), 2016, pp. 1-4: IEEE.

[14] O. Z. Alsulami, M. O. Musa, M. T. Alresheedi, and J. M. H. Elmirghani, "Visible light optical data centre links," in 2019 21st International Conference on Transparent Optical Networks (ICTON), 2019, pp. 1-5: IEEE.

[15] A. E. Eltraify, M. O. Musa, A. Al-Quzweeni, and J. M. H. Elmirghani, "Experimental evaluation of passive optical network based data centre architecture," in 2018 20th International Conference on Transparent Optical Networks (ICTON), 2018, pp. 1-4: IEEE.

[16] A. E. Eltraify, M. O. Musa, A. Al-Quzweeni, and J. M. H. Elmirghani, "Experimental Evaluation of Server Centric Passive Optical Network Based Data Centre Architecture," in 2019 21st International Conference on Transparent Optical Networks (ICTON), 2019, pp. 1-5: IEEE.

[17] J. M. H. Elmirghani, E.-G. Taisir, and A. Hammadi, "Passive optical-based data center networks," ed: Google Patents, 2019.

[18] W. Xia, P. Zhao, Y. Wen, and H. Xie, "A survey on data center networking (DCN): Infrastructure and operations," IEEE communications surveys & tutorials, vol. 19, no. 1, pp. 640-656, 2016.

[19] K. Bilal et al., "A taxonomy and survey on green data center networks," Future Generation Computer Systems, vol. 36, pp. 189-208, 2014.

[20] L. Popa, S. Ratnasamy, G. Iannaccone, A. Krishnamurthy, and I. Stoica, "A cost comparison of datacenter network architectures," in Proceedings of the 6th International COnference, 2010, pp. 1-12.

[21] C. Kachris and I. Tomkos, "A survey on optical interconnects for data centers," IEEE Communications Surveys & Tutorials, vol. 14, no. 4, pp. 1021-1036, 2012.

[22] H. Liu et al., "Circuit switching under the radar with REACToR," in 11th {USENIX} Symposium on Networked Systems Design and Implementation ({NSDI} 14), 2014, pp. 1-15.

[23] C. Kachris and I. Tomkos, "Power consumption evaluation of hybrid WDM PON networks for data centers," in 2011 16th European Conference on Networks and Optical Communications, 2011, pp. 118-121: IEEE.

[24] P. N. Ji, D. Qian, K. Kanonakis, C. Kachris, and I. J. I. J. o. S. T. i. Q. E. Tomkos, "Design and evaluation of a flexible-bandwidth OFDM-based intra-data center interconnect," vol. 19, no. 2, pp. 3700310-3700310, 2012.

[25] K. Wang et al., "ADON: a scalable AWG-based topology for datacenter optical network," vol. 47, no. 8, pp. 2541-2554, 2015.

[26] J. M. H. Elmirghani, T. El-Gorashi, and A. Hammadi, "Passive optical-based data center networks," ed: Leeds, 2015.

[27] A. Hammadi, T. E. El-Gorashi, and J. M. H. Elmirghani, "Energy-efficient software-defined AWGR-based PON data center network," in 2016 18th International Conference on Transparent Optical Networks (ICTON), 2016, pp. 1-5: IEEE.

[28] X. Hong, Y. Yang, Y. Gong, J. J. J. o. O. C. Chen, and Networking, "Passive optical interconnects based on cascading wavelength routing devices for datacenters: A cross-layer perspective," vol. 9, no. 4, pp. C45-C53, 2017.

[29] Y. Hong et al., "A Multi-Floor Arrayed Waveguide Grating Based Architecture With Grid Topology for Datacenter Networks," vol. 8, pp. 107134-107145, 2020.

[30] A. S. Thyagaturu, A. Mercian, M. P. McGarry, M. Reisslein, and W. Kellerer, "Software defined optical networks (SDONs): A comprehensive survey," IEEE Communications Surveys & Tutorials, vol. 18, no. 4, pp. 2738-2786, 2016.

[31] R. Gu, Y. Ji, P. Wei, and S. Zhang, "Software defined flexible and efficient passive optical networks for intra-datacenter communications," Optical Switching and Networking, vol. 14, pp. 289-302, 2014.

[32] B. A. Yosuf, M. Musa, T. Elgorashi, and J. M. H. Elmirghani, "Energy efficient distributed processing for IoT," IEEE Access, vol. 8, pp. 161080-161108, 2020.

[33] S. H. Mohamed, M. B. A. Halim, T. E. Elgorashi, and J. M. H. Elmirghani, "Fog-assisted caching employing solar renewable energy and energy storage devices for video on demand services," vol. 8, pp. 115754-115766, 2020.

[34] Z. T. Al-Azez, A. Q. Lawey, T. E. El-Gorashi, and J. M. H. Elmirghani, "Energy efficient IoT virtualization framework with peer to peer networking and processing," vol. 7, pp. 50697-50709, 2019.

[35] H. A. Alharbi, T. E. Elgorashi, and J. M. H. Elmirghani, "Energy efficient virtual machines placement over cloud-fog network architecture," vol. 8, pp. 94697-94718, 2020.

[36] A. M. Al-Salim, A. Q. Lawey, T. E. El-Gorashi, J. M. H. Elmirghani, and S. Management, "Energy efficient big data networks: Impact of volume and variety," vol. 15, no. 1, pp. 458-474, 2017.

[37] A. M. Al-Salim, T. E. El-Gorashi, A. Q. Lawey, and J. M. H. Elmirghani, "Greening big data networks: Velocity impact," vol. 12, no. 3, pp. 126-135, 2018.

[38] H. M. M. Ali, T. E. El-Gorashi, A. Q. Lawey, and J. M. H. Elmirghani, "Future energy efficient data centers with disaggregated servers," vol. 35, no. 24, pp. 5361-5380, 2017.

[39] N. I. Osman, T. El-Gorashi, L. Krug, and J. M. H. Elmirghani, "Energy-efficient future high-definition TV," vol. 32, no. 13, pp. 2364-2381, 2014.

[40] A. Q. Lawey, T. E. El-Gorashi, and J. M. H. Elmirghani, "BitTorrent content distribution in optical networks," vol. 32, no. 21, pp. 4209-4225, 2014.

[41] M. O. Musa, T. E. El-Gorashi, and J. M. H. Elmirghani, "Bounds on GreenTouch GreenMeter network energy efficiency," vol. 36, no. 23, pp. 5395-5405, 2018.

[42] M. S. Hadi, A. Q. Lawey, T. E. El-Gorashi, and J. M. Elmirghani, "Patient-centric HetNets powered by machine learning and big data analytics for 6G networks," vol. 8, pp. 85639-85655, 2020.

[43] M. Hadi, A. Lawey, T. El-Gorashi, and J. M. H. Elmirghani, "Using machine learning and big data analytics to prioritize outpatients in HetNets," in IEEE INFOCOM 2019-IEEE Conference on Computer Communications Workshops (INFOCOM WKSHPS), 2019, pp. 726-731: IEEE.

[44] M. S. Hadi, A. Q. Lawey, T. E. El-Gorashi, and J. M. H. Elmirghani, "Patient-centric cellular networks optimization using big data analytics," vol. 7, pp. 49279-49296, 2019.

[45] I. S. B. M. Isa, T. E. El-Gorashi, M. O. Musa, and J. M. H. Elmirghani, "Energy Efficient Fog-Based Healthcare Monitoring Infrastructure," vol. 8, pp. 197828-197852, 2020.

[46] M. Musa, T. Elgorashi, J. Elmirghani, and Networking, "Bounds for energy-efficient survivable IP over WDM networks with network coding," vol. 10, no. 5, pp. 471-481, 2018.

[47] M. Musa, T. Elgorashi, J. M. H. Elmirghani, and Networking, "Energy efficient survivable IP-over-WDM networks with network coding," vol. 9, no. 3, pp. 207-217, 2017.

[48] R. Ramaswami, K. Sivarajan, and G. Sasaki, Optical networks: a practical perspective. Morgan Kaufmann, 2009.

[49] A. A. Hammadi, "Future PON data centre networks," University of Leeds, 2016.

[50] A. Hammadi, M. Musa, T. E. El-Gorashi, and J. M. H. Elmirghani, "Resource provisioning for cloud PON AWGR-based data center architecture," in 2016 21st European Conference on Networks and Optical Communications (NOC), 2016, pp. 178-182: IEEE.

[51] C. Clos, "A study of non‐blocking switching networks," Bell System Technical Journal, vol. 32, no. 2, pp. 406-424, 1953.

[52] Cisco. (6 April 2021). Cisco Catalyst 3850 Series Switches. Available: https://www.cisco.com/c/en_in/products/collateral/switches/catalyst-3850-series-switches/datasheet_c78-720918.html

[53] L. Gyarmati and T. A. Trinh, "How can architecture help to reduce energy consumption in data center networking?," in Proceedings of the 1st International Conference on Energy-Efficient Computing and Networking, 2010, pp. 183-186.

[54] W. Hajduczenia, "Power saving techniques in access networks," 00500:: Universidade de Coimbra, 2013.

[55] K. Grobe, M. Roppelt, A. Autenrieth, J.-P. Elbers, and M. J. I. C. M. Eiselt, "Cost and energy consumption analysis of advanced WDM-PONs," vol. 49, no. 2, pp. s25-s32, 2011.